# Two massive rocky planets transiting a K-dwarf 6.5 parsecs away

Michaël Gillon[1], Brice-Olivier Demory[2], Valérie Van Grootel[1], Fatemeh Motalebi[3], Christophe Lovis[3], Andrew Collier Cameron[4], David Charbonneau[5], David Latham[5], Emilio Molinari[6,7], Francesco A. Pepe[3], Damien Ségransan[3], Dimitar Sasselov[5], Stéphane Udry[3], Michel Mayor[3], Giuseppina Micela[8], Giampaolo Piotto[9,10], and Alessandro Sozzetti[11]

[1]Space sciences, Technologies and Astrophysics Research (STAR) Institute, Université de Liège, Allée du 6 Août 19C, Bat. B5C, 4000 Liège, Belgium

[2]Cavendish Laboratory, J J Thomson Avenue, Cambridge CB3 0HE, UK

[3]Observatoire de Genève, Université de Genève, 51 ch. des Maillettes, CH-1290 Sauverny, Switzerland

[4]Center for Exoplanet Science, SUPA, School of Physics & Astronomy, University of St. Andrews, North Haugh, St. Andrews Fife, KY16 9SS, UK

[5]Harvard-Smithsonian Center for Astrophysics, 60 Garden Street, Cambridge, Massachusetts 02138, USA

[6]INAF - Fundación Galileo Galilei, Rambla José Ana Fernandez Pérez 7, 38712 Berña Baja, Spain

[7]INAF - IASF Milano, via Bassini 15, 20133, Milano, Italy

[8]INAF - Osservatorio Astronomico di Palermo, Piazza del Parlamento 1, 90134 Palermo, Italy

[9]Dipartimento di Fisica e Astronomia G. Galilei, Universitá di Padova, Vicolo dell'Osservatorio 2, 35122, Padova, Italy

[10]INAF - Osservatorio Astronomico di Padova, Vicolo dell'Osservatorio 5, 35122, Padova, Italy

[11]INAF - Osservatorio Astrofisico di Torino, via Osservatorio 20, 10025 Pino Torinese, Italy

**HD 219134 is a K-dwarf star at a distance of 6.5 parsecs around which several low-mass planets were recently discovered[1,2]. The Spitzer space telescope detected a transit of the innermost of these planets, HD 219134 b, whose mass and radius (4.5 $M_{Earth}$ and 1.6 $R_{Earth}$ respectively) are consistent with a rocky composition[1]. Here, we report new high-precision time-series photometry of the star acquired with Spitzer revealing that the second innermost planet of the system, HD 219134 c, is also transiting. A global analysis of the Spitzer transit light curves and the most up-to-date HARPS-N velocity data set yields mass and radius estimations of 4.74 ± 0.19 $M_{Earth}$ and 1.602 ± 0.055 $R_{Earth}$ for 219134 b, and of 4.36 ± 0.22 $M_{Earth}$ and 1.511 ± 0.047 $R_{Earth}$ for HD 219134 c. These values suggest rocky compositions for both planets. Thanks to the proximity and the small size of their host star (0.778 ± 0.005 $R_{\odot}$[3]), these two transiting exoplanets - the nearest to the Earth to date - are well-suited for a detailed characterization (precision of a few percent on mass and radius, constraints on the atmospheric properties...) that could give important constraints on the nature and formation mechanism of the ubiquitous short-period planets of a few Earth masses.**

The detection of a transit of HD 219134 b[1], combined with the assumption that the system's planets originated in a common protoplanetary disk, translated into significantly improved transit probabilities for the other planets orbiting the star, especially for HD 219134 c. Using a Monte-Carlo approach[4], and assuming a standard deviation for the orbital inclinations of the system's planets of 2.2 deg, the corresponding value for the solar system, we computed an *a posteriori* transit probability of 21 % for HD 219134 c, significantly greater than its *a priori* geometric transit probability of 5.4 %. We therefore intensified our radial velocity (RV) monitoring of HD 219134 with the HARPS-N spectrograph[5] in the second semester of 2015. Our analysis of the extended HARPS-N dataset (see Methods) resulted in an updated transit ephemeris for HD 219134 c that we used to schedule a monitoring campaign of its transit window with the Spitzer space telescope in spring 2016.

We observed HD 219134 at 4.5µm with the Spitzer/IRAC detector[6] in subarray mode (32x32 pixels windowing of the detector) with an exposure time of 0.01s. The observations were done without dithering, in the so-called PCRS peak-up mode[7]. This maximizes the accuracy in the position of the target on the detector, and thus minimizes the well documented 'pixel phase' effect of IRAC InSb arrays[8]. The observations were performed on 26 Mar and 29 Apr 2016. Both runs lasted 7.5 hrs. They covered, respectively, the first and second part (with significant overlap to avoid any ambiguity due to a partial transit) of the 2σ transit window of HD 219134 c. We also observed the star for 6.5 hours on 16 Mar 2016. These preliminary observations targeted a transit of HD 219134 b in order to confirm the transit of the planet and to improve the precision of its transit ephemeris and radius.

We calibrated the Spitzer images with the computing pipeline S19.2.0 and we performed aperture photometry of the stellar fluxes, using the same aperture of 2.3 pixels for the three datasets. As in ref. 1, the resulting light curves were binned to a time sampling of 30s for the sake of computational speed of the data analysis, but we checked with a short analysis of the original light curves that our results are insensitive to the applied binning.

We used a Markov Chain Monte-Carlo (MCMC) code[9] to explore for each photometric time-series a large range of models, each including an instrumental model to represent Spitzer systematics (see Methods) and assuming or not a transit represented by the eclipse model of ref. 10. We compared the different models using the BIC[11] (Bayesian Information Criterion) as a proxy for the marginal likelihood of the models tested. The presence of a transit was decisively favored (Bayes factor[12] > 1000) for all three time-series (Fig. 1, Table Supplementary 1). This confirmed the transiting nature of HD 219134 b and revealed that of HD 219134 c. After selection of the most likely models, we performed individual MCMC analysis of each transit light curve, including the initial HD 219134 b transit light curve[1], to obtain consistent transit parameters for the two transits of planet b (14 Apr 2015 and 16 Mar 2016) and of planet c (26 Mar and 29 Apr 2016).

We determined a stellar mass of 0.81 ± 0.03 $M_{\odot}$ with the stellar evolution modeling code CLES[13], using as inputs the radius and effective temperature as measured in ref. 3, and the metallicity as derived from spectroscopic analysis[1] (see Table 1). We varied the internal physics for convection efficiency, possible core extra-mixing, and initial helium abundance. The error budget includes the associated uncertainties on the input parameters, but is dominated by the uncertainty on the initial helium abundance when modeling stellar evolution. The uncertainties on convection and extra-mixing parameters have relatively low contributions. Only old stellar ages were obtained (11.0 ± 2.2 Gyr), consistently with the long magnetic cycle and the slow rotation inferred for this star[14]. This old age is also consistent with previous works that favored an age between 6 and 11 Gyr[15,16,17]. Compared to this broad age range, our smaller uncertainty can be attributed to the highly precise stellar radius and temperature constrained by interferometry[3] (Table 1) that we used as inputs to our stellar evolution modeling, unlike these previous works.

We then performed a global MCMC analysis of all our data (HARPS-N RVs + Spitzer photometry, including the initial HD 219134 b transit light curve[1]), to get the strongest constraints on the parameters of the short-period planets orbiting HD 219134 (see Methods). A circular orbit was assumed for HD 219134 b, its proximity to the star resulting in a computed tidal circularization timescale[18] of 80 Myr when assuming a tidal quality factor[19] of 100, corresponding to the maximum value derived for terrestrial planets and satellites of the solar system[19]. The same computation for planet c resulted in a tidal circularization timescale of 2.5 Gyr, so we conservatively left its orbital eccentricity free in our analysis. Table 1 presents the resulting values and error bars for the system parameters, while Figure 1 shows the light curves corrected for the systematics and the best-fit transit models.

As HD 219134 is a well-characterized, bright and nearby star, the detection of the transits of its two inner planets makes possible the first detailed characterization and comparative study of two massive rocky planets orbiting the same star. Notably, an intense RV and photometric follow-up could improve the precision on the planets' masses and radii down to the 3% and 1% levels (currently: 4.5% and 3%), respectively, thanks to very well-constrained values of the stellar mass and radius (see Methods). Assuming rocky compositions for both planets, which is consistent with our measurements (Fig. 2, ref. 20), and applying a semi-empirical mass-radius relation based on Earth's seismic model[21], we infer core mass fractions (CMF) of $0.09^{+0.16}_{-0.09}$ and 0.26±0.17 for planets b and c, respectively. These CMF values have to be compared to a CMF of 0.33 for Earth[21]. At this stage, we can thus only conclude that our current dataset marginally favors a CMF smaller than Earth's for planet b. With the improved precisions on the planets masses and radii mentioned above, the errors on the planets' CMF would drop to 5-6%, making possible much stronger inferences on their compositions. Still, these inferences would rely on the assumptions that both planets have negligible volatile contents and atmospheric extents, as larger CMFs combined with significant volatile contents and/or extended hydrogen-dominated atmospheres could result in the same measured masses and radii[20,22]. Fortunately, the host star is small and bright enough to make it possible to constrain the atmospheric extents and compositions of the planets by transit transmission spectroscopy with the Hubble Space Telescope (HST), and, possibly, by occultation emission spectroscopy with the James Webb Space Telescope (JWST) which is due to launch in 2018 (see Methods).

The theories of formation of short-period planets of a few Earth masses fall into two main classes, one assuming a formation far from the star, outside the snow line, followed by a significant inwards migration by gravitational interaction with the gas disc[23], and the other assuming in-situ formation[24,25]. The latter requires the establishment of a very high surface density of dust grains in the inner protoplanetary disc. These grains then coagulate to form ~cm-sized "pebbles". These are caught by gas drag and migrate inwards to the inner edge of the gas disc where they accumulate, and eventually form close-in planets by gravitational instability or core accretion[26,27]. The two classes of models predict different planetary compositions, the former and latter favoring, respectively, volatile-rich and volatile-poor compositions[28]. Very strong constraints on the planets' compositions could thus help to discriminate their origins. Still, the large irradiation received by the planets during the ~11 Gyrs since their formations could make this discrimination a challenging task even in that case, as it could have significantly altered their initial structures and compositions.

The transiting nature of both HD 219134 b and c increases the probability that planets d and f transit too. Using the formalism of ref. 4, we compute posterior transit probabilities of 13.1 % and 8.1 % for planet f and d, respectively, significantly greater than their prior transit probabilities of 2.5 and 1.5 % (see Table 1). A transit detection for one or both of these planets would increase further the importance of the system for comparative exoplanetology, and a search for their transits is thus highly desirable. Although such a transit search is probably out of reach of ground-based telescopes, it could be performed again by Spitzer, whose operations have been extended to end-2018, or by the space missions TESS[29] and CHEOPS[30] which are due to launch in 2018.

**Author Information.** The authors declare no competing financial interests. Correspondence and requests for materials should be addressed to M.G. (michael.gillon@ulg.ac.be).

**Acknowledgments.** This work is based in part on observations made with the Spitzer Space Telescope, which is operated by the Jet Propulsion Laboratory, California Institute of Technology under a contract with NASA. Support for this work was provided by NASA. M. Gillon is Research Associate at the Belgian Scientific Research Fund (F.R.S-FNRS), and he is extremely grateful to NASA and SSC Director for having supported his searches for RV planets with Spitzer. The research leading to these results has received funding from the ARC grant for Concerted Research Actions, financed by the Wallonia-Brussels Federation. The authors thank N. Lewis for valuable information on the potential for atmospheric characterization of HD 219134 b and c with JWST.

**Author Contributions.** M. G. led the HD219134b+c transit search with Spitzer, planned and analyzed the Spitzer observations, performed the global analysis of the Spitzer and HARPS-N data, and wrote most of the manuscript. B.-O. D. performed an independent analysis of the Spitzer data to verify M. G.'s results. V. V. G. performed the stellar evolutionary modeling of the host star. A. C. C., D. C., D. L., C. L., E. M., F. M., F. A. P., D. S., D. Sé. and S. U. form the HARPS-N science team which managed the RV monitoring of the system. All co-authors have contributed to the writing of the manuscript.

**Online Content.** Methods, along with any additional Supplementary Information display items, are available in the online version of the paper; references unique to these sections appear only in the online paper.

**Table 1 | Parameters of HD 219134 and its four inner planets**

| Parameters | Value | | | |
|---|---|---|---|---|
| **Star** | **HD 219134** | | | |
| Mass ($M_{\odot}$)[a] | 0.81 ± 0.03 | | | |
| Radius ($R_{\odot}$)[b] | 0.778 ± 0.005 | | | |
| Effective temperature (K)[b] | 4699 ± 16 K | | | |
| [Fe/H] (dex.)[c] | +0.11 ± 0.04 | | | |
| Age (Gyr)[a] | 11.0 ± 2.2 | | | |
| Density ($\rho_{\odot}$) | 1.729 ± 0.073 | | | |
| Log. surface gravity (dex.) | 4.567 ± 0.018 | | | |
| Luminosity ($L_{\odot}$) | 0.2646 ± 0.0050 | | | |
| **Planets** | **b** | **c** | **f** | **d** |
| Mid-transit time ($BJD_{TDB}$ - 2450000) | 7126.69913 ± 0.00087 | 7474.04591±0.00088 | 7716.31± 0.50 | 7726.03± 0.63 |
| Orbital period (d) | 3.092926 ± 0.000010 | 6.76458 ± 0.00033 | 22.717 ± 0.015 | 46.859 ± 0.028 |
| Semi-major axis (au) | 0.03876 ± 0.00047 | 0.06530 ± 0.00080 | 0.1463 ± 0.0018 | 0.2370 ± 0.0030 |
| Irradiation ($S_{Earth}$) | 176.2 ± 5.5 | 62.1 ± 1.9 | 12.38 ± 0.38 | 4.72 ± 0.15 |
| Equilibrium temperature (K)[d] | 1015 ± 8 | 782 ± 6 | 522.6 ± 4.1 | 410.5 ± 3.2 |
| Transit depth (ppm) | 358 ± 24 | 318 ± 19 | > 236 ± 8 | > 358 ± 7 |
| Transit impact parameter ($R_{\star}$) | 0.9245 ± 0.0060 | 0.813 ± 0.025 | ? | ? |
| Transit duration (min)[e] | 56.7 ± 1.6 | 99.7 ± 2.6 | < 229 ± 13 | < 337 ± 10 |
| Orbital inclination (deg) | 85.05 ± 0.09 | 87.28 ± 0.10 | ? | ? |
| Orbital eccentricity | 0 (fixed) | 0.062 ± 0.039 | 0.148 ± 0.047 | 0.138 ± 0.025 |
| Argument of pericenter (deg) | 0 (fixed) | $70^{+24}_{-35}$ | 81 ± 17 | 173 ± 10 |
| RV semi-amplitude ($m.s^{-1}$) | 2.381 ± 0.075 | 1.697 ± 0.075 | 1.92 ± 0.10 | 3.34 ± 0.11 |
| Mass ($M_{Earth}$) | 4.74 ± 0.19 | 4.36 ± 0.22 | > 7.30 ± 0.40 | > 16.17 ± 0.64 |
| Radius ($R_{Earth}$)[f] | 1.602 ± 0.055 | 1.511 ± 0.047 | > 1.31 ± 0.02 | > 1.61 ± 0.02 |
| Density ($\rho_{Earth}$) | 1.15 ± 0.13 | 1.26 ± 0.14 | ? | ? |
| Posterior $P_{transit}$ (%)[g] | 100 | 100 | 13.1 | 8.5 |

The stellar mass was derived by stellar evolution modeling, while the planetary parameters result from a combined MCMC analysis of the most recent HARPS-N RV dataset and of the Spitzer transit photometry, including the three new transit light curves presented here.

[a]from our stellar evolution modeling. [b]from ref. 3 [c]from ref. 1. [d]assuming a null Bond albedo. [e]upper limit for f and d (edge-on orbit). [f]lower limit for planets f and d (radius computed with the model of ref. 20 for a pure iron composition). [g]computed by Monte-Carlo simulations using the formalism of ref. 4

**Figure 1 | Spitzer transit photometry of the planets HD 219134 b and c**. Spitzer/IRAC 4.5μm time-series photometry for HD219134, corrected for the instrumental effects, unbinned (cyan dots) and binned per 7.2 min = 0.01d (black circles with error bars, each error bar being the standard deviation for the bin). For each light curve, the best-fit transit model is superimposed in red. The first and second panel show the transits of, respectively, HD 219134 b and HD 219134 c. The photometry is folded on the orbital period of the planets (0 = inferior conjunction).

**Figure 2 | Mass-radius relationship for small planets with precisions on the masses better than 20%.** The solid lines are theoretical mass-radius curves from ref. 20.

## METHODS

### HARPS-North radial velocities

The determination of the transit ephemerides of HD 219134 b and c (see below) was based on 553 radial velocity measurements gathered by the HARPS-N spectrograph[5] between 9 Aug 2012 and 3 Sep 2015, including the 481 measurements presented in ref. 1. We acquired 110 additional HARPS-N measurements between 3 Sep and 9 Nov 2015, expanding our RV time-series to 663 measurements that we used as input data in our global analysis aiming to constrain the orbital and physical parameters of the four inner planets of HD 219134. All the HARPS-N measurements were obtained with the same observational and reduction strategy described in ref. 1.

### Transit ephemerides determination

Our determination of the transit ephemerides of HD 219134 b and c was based on a two step analysis of the first 553 HARPS-N radial velocities. First, we used the Systemic Console software[31] to identify the best-fitting planetary solutions (number of planets + orbits), then we used an adaptive Markov Chain Monte-Carlo (MCMC) code[9] to explore each of these solutions in terms of model marginal likelihood and posterior probability distribution functions (PDFs) of the planetary parameters, including the transit ephemerides of planets b and c. Using the BIC as a proxy for the marginal likelihood of the tested models, we elected as our nominal model one star and four planets on Keplerian orbits about their common center of mass (see Table 1). This was added to a baseline model consisting of polynomial functions of time, cross-correlation function[32] parameters (bisector and width), and the log($R'_{HK}$) stellar activity indicator[33] to represent the remaining signal of planetary (long-period planet(s)) and stellar origins (Supplementary Table 2). The four assumed planets are HD 219134 b, c, and d (ref. 1), and the 22.7d-period planet f (ref. 2) that is firmly detected in the extended (relative to ref. 1) HARPS-N dataset. We measured the quadratic difference between the standard deviation of the best-fit residuals and the mean internal error estimate to derive a `jitter' noise[34] of 1.1 m.s$^{-1}$ which was then added in quadrature to the measurement uncertainties. Including the mid-transit time measured for HD 219134 b in ref. 1 as prior, we explored then thoroughly our selected model with five Markov Chains of 100,000 steps, the first 20% of each chain being considered as its burn-in phase[9] and discarded. The resulting mid-transit ephemerides - that we used to plan our Spitzer observations - were $T_b$ = 2457126.7001 (±0.0010) + N x 3.09321 (±0.00038) $BJD_{TDB}$ and $T_c$ = 2457474.22 ± 0.11 + N x 6.7632 (±0.0017) $BJD_{TDB}$ for, respectively, planet b and c.

### Stellar evolution modeling

We used the CLES code[13] to model the evolution of the host star in order to estimate the stellar mass and the age of the system. We used as inputs the stellar effective temperature from ref. 3, and the metallicity as derived from the spectroscopic iron abundance from ref. 1. We also estimated the stellar metallicity from several elemental abundances[35] and found similar results. For the solar reference we used the abundances of ref. 36. Microscopic diffusion (gravitational settling) of elements were included following the prescription of ref. 37. We varied the internal physics for convection efficiency, possible core extra-mixing (due to overshooting during early evolution phases, rotationally induced mixing, etc.), and initial helium abundance. Since the helium atmospheric abundance cannot be directly measured from spectroscopy, we computed evolutionary tracks with various initial helium abundances, for a quasi-primordial value ($Y_0$=0.25), a protosolar value ($Y_0$=0.27), and values obtained from the general trend observed for the chemical evolution of galaxies (up to $Y_0$~0.30).

**Limb-darkening coefficients**

As in ref. 1, we assumed a quadratic law[38] to represent the limb darkening and its impact on the shape of the transit light curves[10]. Values for the two quadratic limb-darkening coefficients $u_1$ and $u_2$ were interpolated at each step of the Markov chains from the tables of ref. 39 for the Spitzer 4.5μm bandpass, based on the step's values for the stellar effective temperature $T_{eff}$, metallicity [Fe/H], microturbulence speed $\xi_t$, and surface gravity logarithm log $g$. The marginal posterior PDFs of $u_1$ and $u_2$ have as median ± standard deviation, respectively 0.0812 ± 0.0005 and 0.1498 ± 0.0013.

**Individual analysis of the Spitzer light curves**

For each of the four Spitzer photometric time-series, we used the adaptive MCMC code presented in ref. 9 (and references therein) to explore a large range of models, each consisting of a baseline model aiming to represent the photometric variations of instrumental origins, added - or not - to the eclipse model of ref. 10 to represent the transit of planet b (light curves of 14 Apr 2015 and 15 Mar 2016) or c (light curves of 26 Mar and 29 Apr 2019). The tested baseline models (see ref. 1 and references therein for details) consisted of polynomial functions of different external parameters (time, width and position of the stellar image in the Spitzer images, logarithm of time, see Supplementary Table 1), multiplied - or not - by a numerical position model computed at each step of the Markov Chain with the Bliss-Mapping method presented in ref. 40. For each light curve and for each model, we made a short MCMC analysis (one Markov chain of 10,000 steps), and we used the BIC[11] (Bayesian Information Criterion) of the best-fit solution as a proxy for the model marginal likelihood. Furthermore, for each light curve and for each tested baseline model, we computed the BIC difference between the models without and with transit, $BIC_{nt} - BIC_t$, to estimate the Bayes factor[12] in favour of the transit hypothesis computed as $e^{(BIC_{nt}-BIC_t)/2}$.

For the three light curves measured in 2016, the baseline models selected by minimization of the BIC comprised the sum of quadratic functions of the x- and y-positions of the stellar images and of linear functions of the full widths at half-maximum of the stellar images in the x- and y-directions, multiplied by a Bliss-Mapping sub-pixel model. For the first light curve, measured in 2015, the polynomial function included also a linear function of time and a quadratic function of the logarithm of time required to represent a sharp decrease of the detector response at the beginning of the observations[1]. The presence of a transit was decisively favored (Bayes factor > 1000) for all four time-series and for all the tested baseline models (Supplementary Table 1).

For each light curve, we then performed a longer MCMC analysis comprising five Markov chains of 100,000 steps, similar in detail to the global analysis described below. These individual analyses resulted in consistent transit parameters for the two transits of planet b (14 Apr 2015[1] and 16 Mar 2016) and of planet c (26 Mar and 29 Apr 2016). The selection of a model with transit for the four light curves and the consistency of the fitted transit parameters with the ones expected for the transits of the planets b and c detected by RV[1,2] allowed us to firmly conclude to the transiting nature of HD 219134 c and to confirm the one of HD 219134 b.

## Global MCMC analysis

We performed a global MCMC analysis of the HARPS-N and Spitzer time-series to derive the strongest possible constraints on the parameters of the system. This global analysis consisted in five Markov chains of 100,000 steps. Their convergence was successfully checked with the statistical test of ref. 41. Supplementary Table 2 presents the models selected by minimization of the BIC for each individual data set. The model assumed to represent the RVs was the same as for our transit ephemerides determination (see above).

Supplementary Table 2 also shows for each light curve the error correction factor (CF) representing both the over- or under-estimation of the white noise of each measurement and the presence of correlated (red) noise in the data (see ref. 1 for details). The jump parameters of the MCMC, i.e. the parameters randomly perturbed at each step of the Markov chains, were the following.

- The stellar mass, radius, effective temperature, and metallicity [Fe/H]. For these four parameters, normal prior PDFs based on the values given in Table 1 were assumed.

- For HD 219134 b and c, the planet/star area ratio $dF = (R_p/R_*)^2$, and the parameter $b' = a \cos i / R_*$, where $a$ is the orbital semi-major axis and $i$ is the orbital inclination. $b'$ is the transit impact parameter in the case of circular orbit. For the two other planets, $dF$ was fixed to an arbitrary value and $b'$ was fixed to zero, as no transit of these planets was expected to happen during the Spitzer observations.

- For the four planets, the orbital period $P$, the time of inferior conjunction $T_0$ (corresponding to the mid-transit time for transiting planets), the parameter $K_2 = K\sqrt{1-e^2}P^{1/3}$, where $K$ is the RV semi-amplitude and $e$ is the orbital eccentricity, and - except for HD 219134 b for which a circular orbit was assumed - the two parameters $\sqrt{e}\cos\omega$ and $\sqrt{e}\sin\omega$, where ω is the argument of periastron.

At each step of the MCMC, orbital, physical, and eventually transit (duration, impact parameter) parameters of the planets are computed from their jump parameters (see ref. 12 and references therein).

Table 1 presents for the four planets the resulting median and 1-σ errors of the resulting posterior PDFs for the most physically relevant parameters. Figure 1 shows for the four Spitzer light curves the best-fit transit models and the light curves divided by the best-fit baseline models. Supplementary Figure 1 shows the raw Spitzer light curves and the corresponding best-fit global model (transit + baseline). Supplementary Figure 2 shows the best-fit Keplerian RV models for the four planets.

In a final stage, we performed a global analysis of the Spitzer photometry alone, assuming a uniform prior PDF for the stellar radius to derive the value of the stellar density constrained only by the transit photometry[42]. For this analysis, we assumed the orbits of both planets b and c to be circular. It resulted in a stellar density of $1.7^{+1.3}_{-0.8}\,\rho_{\odot}$, in excellent agreement with the density of 1.719±0.073 $\rho_{\odot}$ derived when using the *a priori* knowledge of the stellar radius. This test brings a further validation of the planetary origin of the transit signals.

**Potential for future improvements in precision of radii and masses of planets b and c**

Transit observations allow measurement of the planet/star area ratio $dF = (R_p/R_*)^2$. The derivative of this formula shows that the relative error on the planetary radius $\sigma R_p/R_p$ is proportional to the relative error on the stellar radius, $\sigma R_*/R_*$, and to half of the relative error on the transit depth, 1/2 $\sigma dF/dF$. Asssuming that $\sigma R_*/R_*$ and $dF/dF$ are uncorrelated, $\sigma R_p/R_p$ can thus be expressed as the quadratic sum:

$$\sigma dF/dF \;=\; \sqrt{(\sigma R_*/R_*)^2 + (\sigma dF/2dF)^2} \qquad (1)$$

Injecting the values and errors of $R_*$ and $dF$ shown in Table 1 for planets b and c into equation (1) results in relative errors of, respectively, 3.4% and 3.1%, in perfect match with the relative errors deduced from the MCMC.

Assuming an improvement of a factor ~5 of the precisions on the planets' transit depths resulting from an intense photometric monitoring campaign of their transits with, e.g., Spitzer (at least 50 transits observed for both planets), the formula above predicts relative errors <1% on the planets' radii, thanks to the precision of 0.64% on the stellar radius.

Assuming a circular orbit and perfectly-determined orbital period and inclination, the same approach shows that the relative error on the mass of a planet deduced from RV measurements can be expressed[43] as the quadratic sum:

$$\sigma M_p/M_p \;=\; \sqrt{(\sigma K/K)^2 + (2\sigma M_*/3M_*)^2} \qquad (2)$$

Injecting the values and errors of $M_*$ and $K$ shown in Table 1 for planets b and c into this equation results in relative errors of 4% and 5% for the masses of planet b and c, respectively, in perfect agreement with the relative errors $\sigma M_p/M_p$ deduced from the MCMC (Table 1). Considering the precision of 3.7% on the stellar mass, equation 2 predicts precisions ~3% for the planets' masses, provided improvements of a factor ~2 for the precisions of the RV semi-amplitudes that could be achieved after several years of an intensive RV monitoring campaign of the star (at least 2000 new RV measurements).

**Potential for atmospheric characterization of the planets with HST and JWST**

HD 219134 is a relatively small star (0.778±0.005 $R_{\odot}$)[3] and is a very bright infrared source (K=3.25). Both features make it an *a priori* favorable target for the atmospheric characterization of its two transit inner planets by eclipse transmission and emission spectroscopy[42]. To quantify this potential, we computed for both planets estimates for the amplitudes of the signals in transmission and emission under different assumptions. For transmission, we used the formula[42]:

$$\Delta dF = 2dFN_H(H/R_p), \qquad (3)$$

where $\Delta dF$ is the increase in transit depth at a wavelength corresponding to a strong atomic or molecular transition, $R_p$ is the planet's radius, $N_H$ is the effective atmospheric extent in atmospheric scale height for a strong transition, and $H$ is the atmospheric scale height given by:

$$H = k_b T/\mu_m g, \quad (4)$$

where $k_b$ is Boltzmann's constant, $T$ is the temperature of the upper atmosphere at the planet's terminator, $\mu_m$ is the mean molecular mass, and $g$ is the planet' surface gravity. For these order of magnitude computations, we equated $T$ to the equilibrium temperature of the planet's dayside $T_{eq}$ computed as:

$$T_{eq} = T_{eff,*} (R_*/a)^{1/2} [f'(1-A_B)]^{1/4}, \quad (5)$$

where $T_{eff,*}$ and $R_*$ are the star's effective temperature and radius, $a$ is the planet semi-major axis, $A_B$ is its Bond albedo (assumed to be 0.25) and $f'$ is a factor indicating the efficiency of the heat distribution from the day-side to the night-side. Given the short orbital periods of the planets, we can assume that they are tidally locked[44], i.e. that tidal forces have trapped them in a 1:1 spin-orbit resonance, like the Moon to the Earth. If such a planet harbors a dense enough atmosphere, it will efficiently distribute the heat to its night side and $f'$ will be close to 1/4. Otherwise, the thermal gradient between both hemispheres will be large and $f'$ could be as large as 2/3. For our transmission computations, we assumed $f'$=1/4.

$N_H$ is a number ranging between 0 and 10, depending on the presence of clouds, on the spectral resolution and on the covered spectral range[22,45]. For these estimates, we assumed $N_H$ = 5. For the atmospheric mean molecular mass $\mu_p$, we tried the following values (in atomic mass units): 2, 18, and 44, corresponding respectively to pure molecular hydrogen, water, and carbon dioxide atmospheres.

For emission estimates, we computed the depth of the occultation of the planets at different infrared wavelengths (1.5-5-10-15-20 microns), using the formula:

$$dFoc,\lambda = dF\, B(T_{eq,p},\lambda)/B(T_{eff,*},\lambda), \quad (6)$$

where $B(T,\lambda)$ is the Planck function at wavelength $\lambda$ for a black body of temperature $T$. For both planets, we computed the occultation depths assuming $f'$=1/4 and $f'$=2/3 in equation (5).

Supplementary Table 3 presents our computed signal estimates for both transmission and emission. The ~5σ detection of a hydrogen-dominated atmosphere by transit transmission spectroscopy would require spectrophotometric precisions of the order of 15 and 10 ppm for, respectively, planet b and c. Spectrophotometric precisions of ~20ppm per transit have been recently reached for the planet 55 Cnc e with the Wide-Field Camera 3 aboard HST, for a spectral sampling ~20 nm and within a spectral range of 1.1 to 1.7 μm[46]. HD 219134 is slightly brighter than 55 Cnc in this spectral range (J=3.9 *vs* J=4.6), but for such bright stars, the precisions are limited not by the photon noise but by the instrumental systematics. It is thus reasonable to assume that HST/WFC3 could reach for HD 219134 precisions of 20 ppm per transit for a similar spectral sampling, which would make possible detecting molecular signatures for both planets, provided that they harbor extended $H_2$-dominated atmospheres. In case of more compact (e.g. $H_2O$- or $CO_2$-dominated) atmospheres, the transmission signals would have amplitudes of only a few ppm, which would be out of reach for HST.

In theory, the largest aperture of JWST should allow it to reach photon noise levels less than 1 ppm in transmission spectroscopy for a star as bright as HD 219134. Unfortunately, the star is *too* bright for unsaturated spectroscopic observations for most JWST instruments[47]. The exception is the NIRCAM instrument, for which a new science mode has been recently introduced that will make it possible to observe spectra in the 1 μm to 2 μm range for stars as bright as *K=2*, or even brighter[48]. Here too, the precisions will be limited not by the photon noise but by correlated noises of instrumental origins. Considering the similitude between the NIRCAM and WFC3 instruments, it is reasonable to assume a spectrophotometric precision limit of ~20ppm per transit, an order of magnitude too large for the detection of compact atmospheres around the planets.

It is possible that the large irradiation received by the planets could produce extended exospheres of hydrogen photoevaporated from their atmospheres or of heavier atomic species carried upward along with the hydrogen flow through collisions or ejected from the planets' surfaces by the strong stellar winds[49]. As such an expanding exosphere can cover a much larger fraction of the star than the planetary disk, transit transmission spectroscopy performed at wavelengths corresponding to strong electronic transitions - in the ultraviolet and the visible - is a valuable method to detect and study them, as demonstrated, e.g., by the recent detection of an hydrogen exosphere around the hot Neptune GJ 436b[50]. Such exosphere detections could be attempted for different species producing particularly strong lines (e.g. H, Na, $Ca^+$, Mg...) with HST and even ground-based telescopes[49], and could place valuable constraints on the planets' atmospheric compositions.

The expected emission signals of both planets, especially of planet b, are large enough beyond 5 μm to make in theory their occultations detectable with current technology (Supplementary Table 3). The JWST/MIRI instrument would be well suited to measure the planets' dayside emission spectra in the 5-29 μm, but for the brightness of the star that would make it saturate the detectors for all MIRI instrumental modes[47], except maybe the MRS mode that has still to be validated for time-series observation (N. Lewis, private communication, ref. 51).

## Code availability

The photometric reduction of the Spitzer images was done with a custom-made IRAF pipeline. IRAF is distributed by the National Optical Astronomy Observatory, which is operated by the Association of Universities for Research in Astronomy, Inc., under cooperative agreement with the National Science Foundation. The MCMC software used to analyse the photometric and RV time-series data is a custom Fortran 90 code. Our Spitzer photometric pipeline and MCMC code are available upon request to the first author.

## Data availability

The data that support the plots within this paper and other findings of this study are available from the corresponding author upon reasonable request.

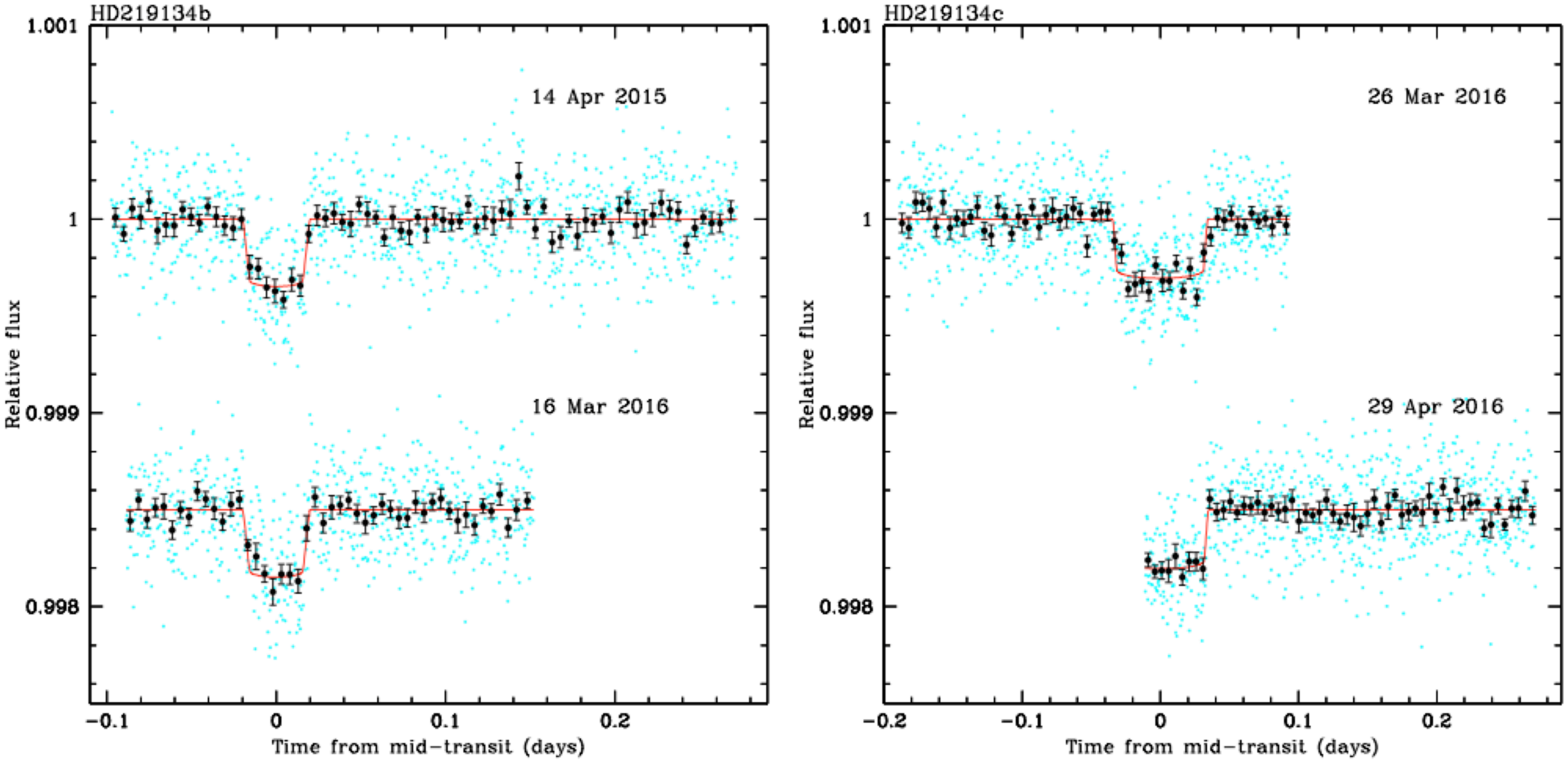

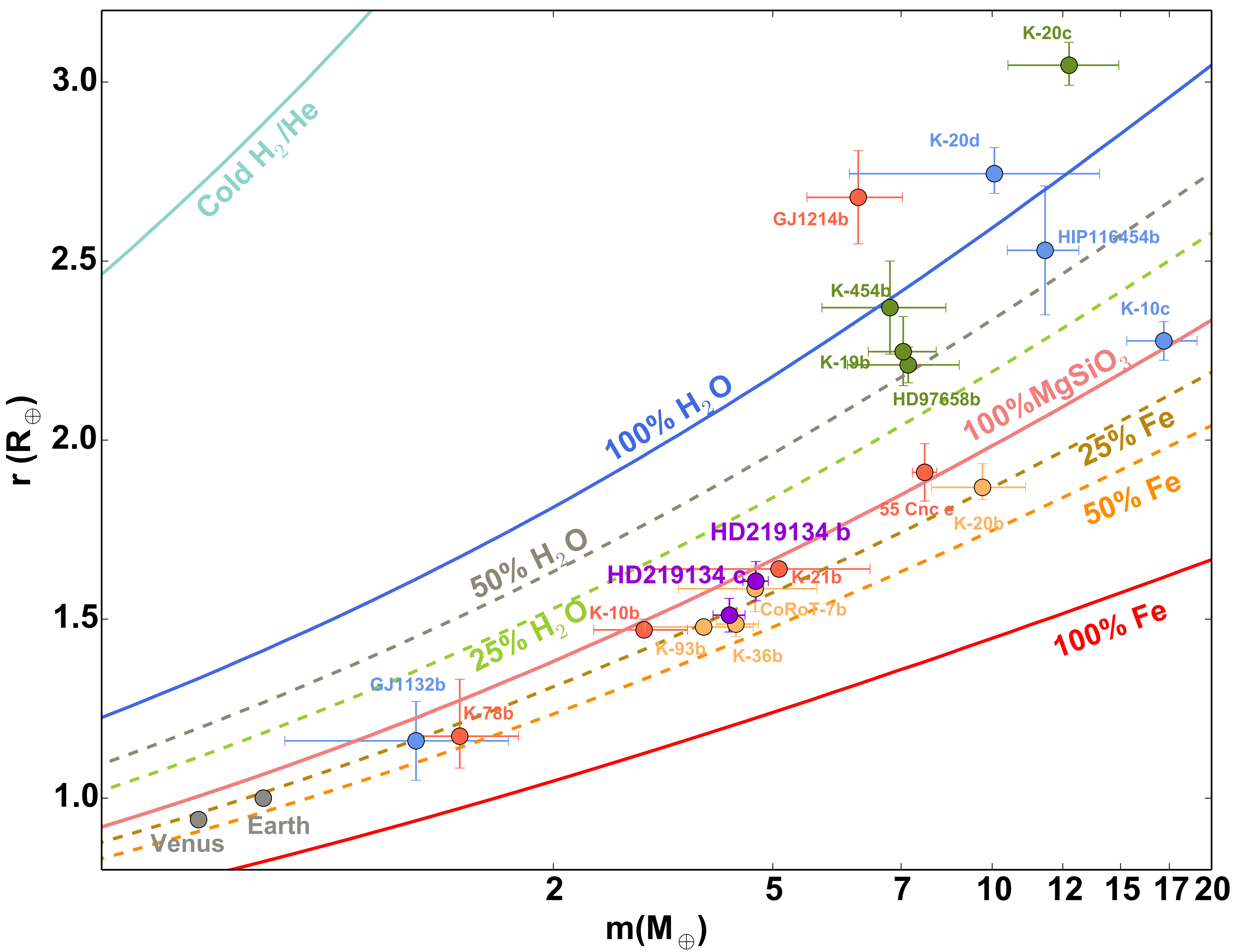

# Two massive rocky planets transiting a K-dwarf 6.5 parsecs away

# Supplementary Information

## 1. Supplementary Figures

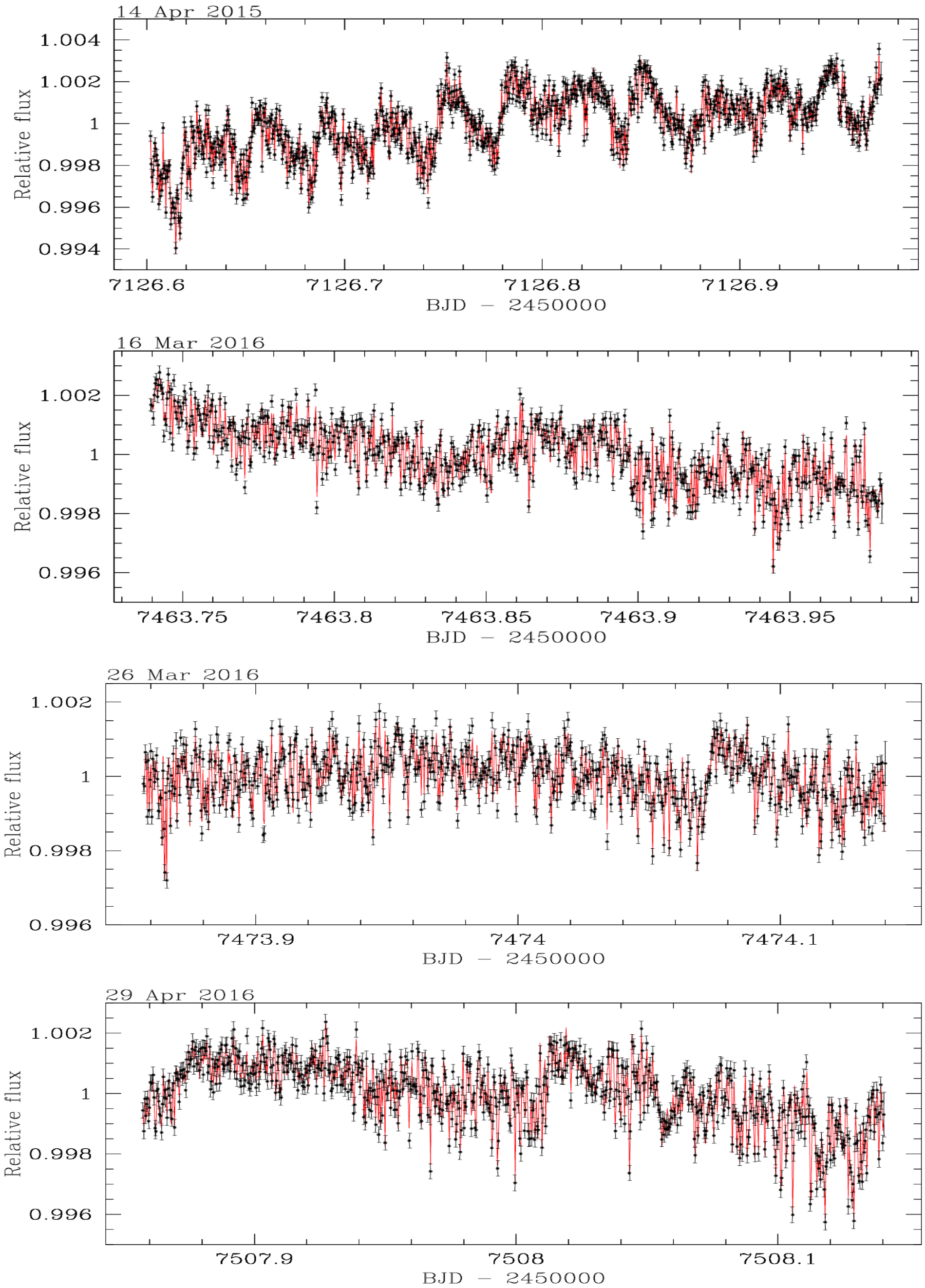

**Supplementary Figure 1 | Raw Spitzer light curves used in this work.** The light curves are shown in chronological order from top to bottom. Each individual measurement correspond to a cube of 64 Spitzer subarray images (see ref. 1 for detail). The measurements and error bars are, respectively, the averages of the 64 subaray flux measurements and their standard errors. The measurement are overplotted on the best-fit transit + baseline models.

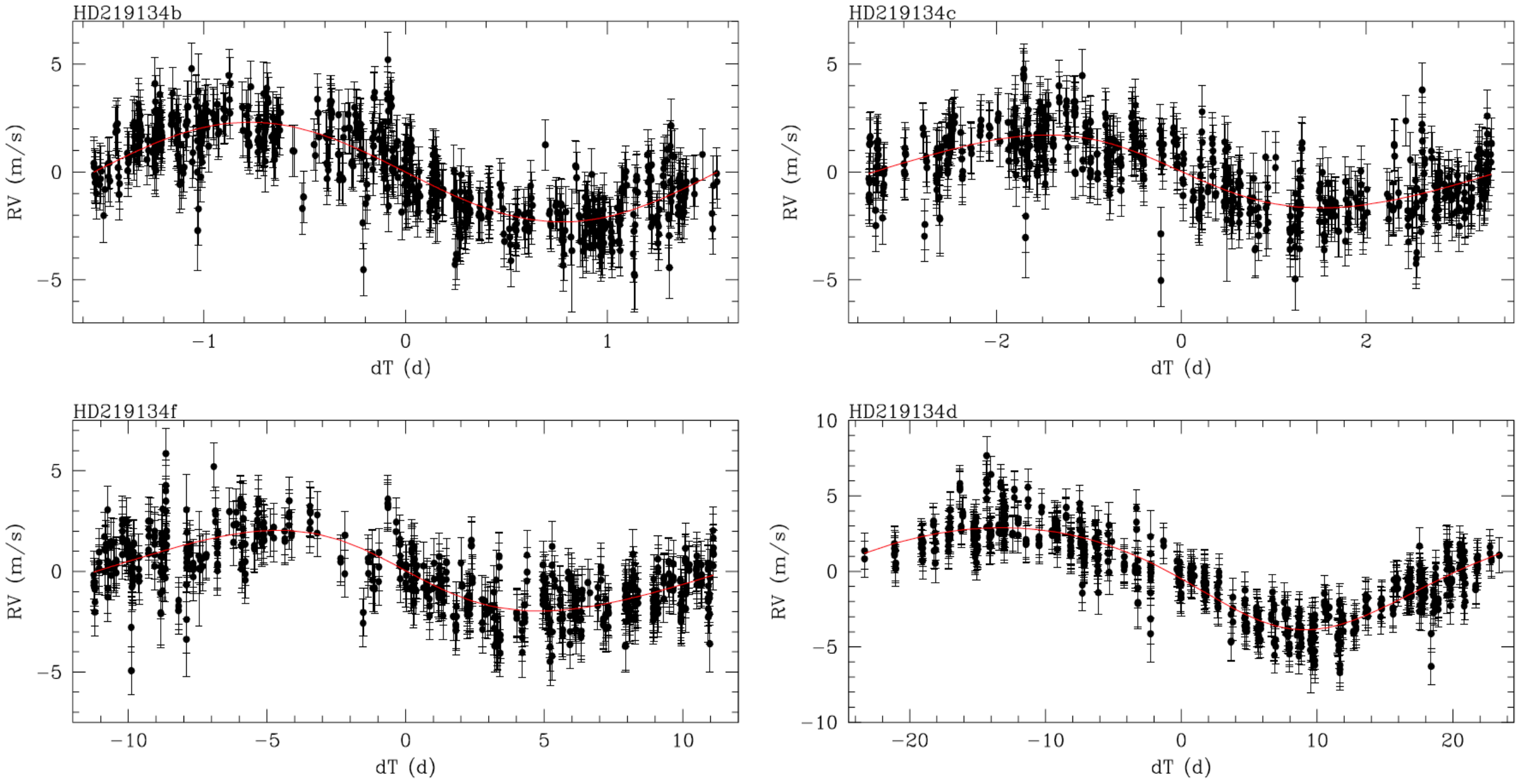

**Supplementary Figure 2 | Time-folded radial velocity measurements of HD 219134.** The measurements are time-folded on the best-fit transit ephemeris (0 = mid-transit) of the four inner planets of the system, after removal of the contributions of the star and the other planets. The corresponding best-fit Keplerian models are shown as red solid lines. The error bars are the quadratic sum of the measurement uncertainties and a 'jitter' noise of 1.1 m.s$^{-1}$ (see Methods).

## 2. Supplementary Tables

| Model | 14 Apr 2015 | 16 Mar 2016 | 26 Mar 2016 | 29 Apr 2016 |
|---|---|---|---|---|
| BM | $BIC_{nt}$ = 4410.4<br>$BIC_t$ = 4265.5<br>$BF_t$ = 2.9 $10^{31}$ | $BIC_{nt}$ = 5159.3<br>$BIC_t$ = 5136.7<br>$BF_t$ = 8.1 $10^{4}$ | $BIC_{nt}$ = 2066.6<br>$BIC_t$ = 1920.3<br>$BF_t$ = 5.9 $10^{31}$ | $BIC_{nt}$ = 2154.6<br>$BIC_t$ = 2089.1<br>$BF_t$ = 1.7 $10^{14}$ |
| BM + p(x,y) | $BIC_{nt}$ = 3517.2<br>$BIC_t$ = 3322.0<br>$BF_t$ = 2.4 $10^{42}$ | $BIC_{nt}$ = 4547.6<br>$BIC_t$ = 4515.7<br>$BF_t$ = 8.4 $10^{6}$ | $BIC_{nt}$ = 1665.9<br>$BIC_t$ = 1524.7<br>$BF_t$ = 4.6 $10^{30}$ | $BIC_{nt}$ = 1841.8<br>$BIC_t$ = 1780.4<br>$BF_t$ = 2.1 $10^{13}$ |
| BM + p($x^2$,$y^2$) | $BIC_{nt}$ = 3367.2<br>$BIC_t$ = 3159.0<br>$BF_t$ = 1.6 $10^{45}$ | $BIC_{nt}$ = 4510.9<br>$BIC_t$ = 4474.2<br>$BF_t$ = 9.3 $10^{7}$ | $BIC_{nt}$ = 1650.1<br>$BIC_t$ = 1513.8<br>$BF_t$ = 4.0 $10^{29}$ | $BIC_{nt}$ = 1596.1<br>$BIC_t$ = 1563.4<br>$BF_t$ = 1.3 $10^{7}$ |
| BM + p($x^2$,$y^2$,$f_x$) | $BIC_{nt}$ = 1601.6<br>$BIC_t$ = 1345.8<br>$BF_t$ = 3.5 $10^{55}$ | $BIC_{nt}$ = 964.3<br>$BIC_t$ = 829.4<br>$BF_t$ = 2.0 $10^{29}$ | $BIC_{nt}$ = 1067.6<br>$BIC_t$ = 855.7<br>$BF_t$ = 1.0 $10^{46}$ | $BIC_{nt}$ = 970.3<br>$BIC_t$ = 895.4<br>$BF_t$ = 1.8 $10^{16}$ |
| BM + p($x^2$,$y^2$,$f_x^2$) | $BIC_{nt}$ = 1605.5<br>$BIC_t$ = 1351.5<br>$BF_t$ = 1.4 $10^{55}$ | $BIC_{nt}$ = 966.4<br>$BIC_t$ = 832.6<br>$BF_t$ = 1.1 $10^{29}$ | $BIC_{nt}$ = 1069.1<br>$BIC_t$ = 863.7<br>$BF_t$ = 4.0 $10^{44}$ | $BIC_{nt}$ = 974.6<br>$BIC_t$ = 897.7<br>$BF_t$ = 5.0 $10^{16}$ |
| BM + p($x^2$,$y^2$,$f_x$,$f_y$) | $BIC_{nt}$ = 1506.3<br>$BIC_t$ = 1269.9<br>$BF_t$ = 2.2 $10^{51}$ | $BIC_{nt}$ = 914.2<br>**$BIC_t$ = 752.6**<br>$BF_t$ = 1.2 $10^{35}$ | $BIC_{nt}$ = 1071.8<br>**$BIC_t$ = 832.6**<br>$BF_t$ = 8.7 $10^{51}$ | $BIC_{nt}$ = 953.1<br>**$BIC_t$ = 877.5**<br>$BF_t$ = 2.6 $10^{16}$ |
| p($x^2$,$y^2$,$f_x$,$f_y$) | $BIC_{nt}$ = 1576.5<br>$BIC_t$ = 1324.3<br>$BF_t$ = 5.8 $10^{54}$ | $BIC_{nt}$ = 952.0<br>$BIC_t$ = 782.2<br>$BF_t$ = 7.4 $10^{36}$ | $BIC_{nt}$ = 1124.6<br>$BIC_t$ = 871.9<br>$BF_t$ = 7.5 $10^{54}$ | $BIC_{nt}$ = 1001.3<br>$BIC_t$ = 909.9<br>$BF_t$ = 7.0 $10^{19}$ |
| BM + p($x^2$,$y^2$,$f_x$,$f_y^2$) | $BIC_{nt}$ = 1510.2<br>$BIC_t$ = 1276.2<br>$BF_t$ = 6.5 $10^{50}$ | $BIC_{nt}$ = 918.4<br>$BIC_t$ = 758.9<br>$BF_t$ = 4.3 $10^{34}$ | $BIC_{nt}$ = 1068.2<br>$BIC_t$ = 837.9<br>$BF_t$ = 1.0 $10^{50}$ | $BIC_{nt}$ = 960.6<br>$BIC_t$ = 881.9<br>$BF_t$ = 1.2 $10^{17}$ |
| BM + p($x^2$,$y^2$,$f_x$,$f_y$,t) | $BIC_{nt}$ = 1510.4<br>$BIC_t$ = 1267.0<br>$BF_t$ = 7.1 $10^{52}$ | $BIC_{nt}$ = 916.7<br>$BIC_t$ = 759.1<br>$BF_t$ = 1.7 $10^{34}$ | $BIC_{nt}$ = 1007.4<br>$BIC_t$ = 835.3<br>$BF_t$ = 2.3 $10^{37}$ | $BIC_{nt}$ = 949.9<br>$BIC_t$ = 884.8<br>$BF_t$ = 1.4 $10^{14}$ |
| BM + p($x^2$,$y^2$,$f_x$,$f_y$,$t^2$) | $BIC_{nt}$ = 1497.1<br>$BIC_t$ = 1268.0<br>$BF_t$ = 5.6 $10^{49}$ | $BIC_{nt}$ = 906.7<br>$BIC_t$ = 764.5<br>$BF_t$ = 7.6 $10^{30}$ | $BIC_{nt}$ = 1013.0<br>$BIC_t$ = 841.1<br>$BF_t$ = 2.1 $10^{37}$ | $BIC_{nt}$ = 936.4<br>$BIC_t$ = 890.2<br>$BF_t$ = 1.1 $10^{10}$ |
| BM + p($x^2$,$y^2$,$f_x$,$f_y$,t,l) | $BIC_{nt}$ = 1406.9<br>$BIC_t$ = 1233.5<br>$BF_t$ = 4.5 $10^{37}$ | $BIC_{nt}$ = 913.1<br>$BIC_t$ = 764.9<br>$BF_t$ = 1.5 $10^{32}$ | $BIC_{nt}$ = 1013.3<br>$BIC_t$ = 841.6<br>$BF_t$ = 1.9 $10^{37}$ | $BIC_{nt}$ = 930.4<br>$BIC_t$ = 889.4<br>$BF_t$ = 8.0 $10^{8}$ |
| BM + p($x^2$,$y^2$,$f_x$,$f_y$,t,$l^2$) | $BIC_{nt}$ = 1377.1<br>**$BIC_t$ = 1207.6**<br>$BF_t$ = 6.4 $10^{36}$ | $BIC_{nt}$ = 909.7<br>$BIC_t$ = 771.3<br>$BF_t$ = 1.1 $10^{30}$ | $BIC_{nt}$ = 1016.0<br>$BIC_t$ = 847.5<br>$BF_t$ = 3.9 $10^{36}$ | $BIC_{nt}$ = 933.3<br>$BIC_t$ = 895.2<br>$BF_t$ = 1.9 $10^{8}$ |

**Supplementary Table 1 | Models comparison for the Spitzer light curves**. For each Spitzer light curve (column 2 to 5) and for each tested baseline model (line 2 to 13), this table shows the BIC computed for the best-fit model assuming ($BIC_t$) or not ($BIC_{nt}$) a transit of planet b (two first light curves) or c (two last light curves). It also shows the Bayes factor in favor of the transit model ($BF_t$) computed as $e^{(BIC_{nt}-BIC_t)/2}$. BM = Bliss-Mapping. p($\alpha^n$) denotes a polynomial function of order n of the parameter α, with α that can be t = time, $f_x$ and $f_y$ = the full widths at half-maximum of the stellar image in the x- and y-directions, x and y = x- and y-positions of the stellar image on the detector, and l = logarithm of time. For each light curve, the lowest BIC - corresponding to the selected model - is shown in red.

| Input data | Number of points | Assumed model | Error correction |
|---|---|---|---|
| HARPS-N RVs | 663 | $p(t^4 + CCF_{bis}^4 + CCF_{width}^3 + \log(R'_{HK})^3$ | $\sigma_{jitter} = 1.1 m.s^{-1}$ |
| Spitzer 14 Apr 2015 | 1044 | $p(x^2,y^2,f_x,f_y,t,l^2)$ + BM | CF = 1.29 |
| Spitzer 16 Mar 2016 | 680 | $p(x^2,y^2,f_x,f_y)$ + BM | CF = 1.09 |
| Spitzer 26 Mar 2016 | 799 | $p(x^2,y^2,f_x,f_y)$ + BM | CF = 1.01 |
| Spitzer 29 Apr 2016 | 799 | $p(x^2,y^2,f_x,f_y)$ + BM | CF = 1.25 |

**Supplementary Table 2 | Data used in the global MCMC analysis.** For each dataset is given the number of measurements, and the assumed model and error correction. $p(\alpha^n)$ denotes a polynomial function of order n of the parameter α, with α that can be t = time, $CCF_{bis}$ and $CCF_{width}$ = the bisector and full-width at half maximum of the CCF of the spectrum, $\log(R'_{HK})$ = the logarithm of the spectroscopic activity index $R'_{HK}$, $f_x$ and $f_y$ = the full widths at half-maximum of the stellar image in the x- and y-directions, x and y = x- and y-positions of the stellar image on the detector, and l = logarithm of time. BM = Bliss-Mapping, CF = Correction Factor. See text and references therein for detail.

| Signal | HD 219134 b | HD 219134 c |
|---|---|---|
| Transmission ($H_2$ atmosphere) | 75 ppm | 55 ppm |
| Transmission ($H_2O$ atmosphere) | 8 ppm | 6 ppm |
| Transmission ($CO_2$ atmosphere) | 3.5 ppm | 2.5 ppm |
| Emission 1.5 μm | 0.1 - 0.8 ppm | 0 - 0.1 ppm |
| Emission 5 μm | 15 - 30 ppm | 5 - 13 ppm |
| Emission 10 μm | 35 - 55 ppm | 18 - 30 ppm |
| Emission 15 μm | 46 - 66 ppm | 26 - 40 ppm |
| Emission 20 μm | 51- 72 ppm | 31 - 45 ppm |

**Supplementary Table 3 | Transmission and emission signal estimates for planets b and c.** Estimates of the signal amplitudes in transit transmission and occultation emission spectroscopy for planets HD 219134 b and c. For the emission estimates, the two values correspond to heat distribution factors *f'* = 1/4 and 2/3 (see Methods).